\documentclass[useAMS,usenatbib]{mn2e}

\usepackage{times,graphicx,amssymb}

\title[Introducing a Hybrid Radiative Transfer Method for Smoothed Particle Hydrodynamics]{Introducing a Hybrid Radiative Transfer Method for Smoothed Particle Hydrodynamics}
\author[Duncan Forgan, Ken Rice, Dimitris Stamatellos and Anthony Whitworth]{Duncan Forgan $^{1}$\thanks{E-mail:
dhf@roe.ac.uk}, Ken Rice$^{1}$, Dimitris Stamatellos$^{2}$ and Anthony Whitworth$^{2}$ \\
$^{1}$Scottish Universities Physics Alliance (SUPA), Institute for Astronomy, University of Edinburgh, Blackford Hill, Edinburgh, EH9 3HJ, Scotland, UK \\
$^{2}$School for Physics and Astronomy, Cardiff University, 5 The Parade, Cardiff, CF24 3AA, Wales, UK}
\begin{document}

\date{Accepted 28th November 2008}

\pagerange{\pageref{firstpage}--\pageref{lastpage}} \pubyear{0000}

\maketitle

\label{firstpage}

\begin{abstract}
A new means of incorporating radiative transfer into smoothed particle hydrodynamics (SPH) is introduced, which builds on the success of two previous methods - the polytropic cooling approximation as devised by \citet{Stam_2007}, and flux limited diffusion (e.g. \citealt{Mayer}).  This hybrid method preserves the strengths of its individual components, while removing the need for atmosphere matching or other boundary conditions to marry optically thick and optically thin regions.  The code uses a non-trivial equation of state to calculate temperatures and opacities of SPH particles, which captures the effects of H$_{2}$ dissociation, H$^{0}$ ionisation, He$^{0}$ and He$^{+}$ ionisation, ice evaporation, dust sublimation, molecular absorption, bound-free and free-free transitions and electron scattering.  The method is tested in several scenarios, including: (1) the evolution of a \(0.07\,M_{\odot}\) protoplanetary disc surrounding a \(0.5\,M_{\odot}\) star; (2) the collapse of a \(1\,M_{\odot}\) protostellar cloud, and (3) the thermal relaxation of temperature fluctuations in a static homogeneous sphere.

\end{abstract}


\section{Introduction}

\noindent Smoothed Particle Hydrodynamics (SPH) \citep{Lucy,Gingold_Monaghan,Monaghan_92} is a Lagrangian method which represents a fluid by a distribution of particles.  Each particle is assigned a mass, position, internal energy and velocity: state variables such as density and pressure can then be calculated by interpolation - see reviews by \citet{Monaghan_92, Monaghan_05}.  The effects of gravitation are also included as standard in most SPH codes.  Recently, many authors have constructed  versions of SPH with effects of radiative transfer \citep{WB_1,Stam_2005_1,Stam_2005_2,Mayer}.  A full description of polychromatic 3D radiative transfer is currently not possible within SPH (at least while current computational limitations prevent it).  Even describing a snapshot from a simulation using full polychromatic radiative transfer is quite expensive \citep{Stam_2005_1,Stam_2005_2}.  In the past, approximations to individual features of radiative transfer were used - for example the cooling time formalism: \(\dot{U} = \frac{-U}{t_{cool}}\), \citep{Ken_1} which only describes energy loss from the system, and does not model transport of energy between neighbouring particles.  \\

\noindent An example of where radiative transfer plays a fundamental role is in the physics of gravitational instabilities (GIs) in protoplanetary discs. The simple parametrisation of the cooling time method allowed these GIs to be probed and characterised effectively.  Gravitational fragmentation of protoplanetary discs is key to the disc instability theory of giant planet formation \citep{Boss_science}.  However, it is disputed whether fragmentation can indeed occur, with strong debate between different groups using different methods of simulation and different formalisms for radiative transfer.  At the time of writing there is no strong consensus as to whether gravitational instability and fragmentation in protoplanetary discs can be a realistic mechanism for giant planet formation.  \\

\noindent For fragmentation to occur, the disc must become gravitationally unstable, so that gravity can overcome pressure support and rotational support.  The disc can become gravitationally unstable to axisymmetric instabilities if \citep{Toomre_1964}:

\begin{equation} Q = \frac{c_s \kappa}{\pi G \Sigma} < 1 \end{equation}

\noindent where \(c_s\) is the local sound speed, \(\kappa\) is the epicyclic frequency, and \(\Sigma\) is the surface density of the disc.  In a Keplerian disc, the epicyclic frequency is replaced by the angular frequency, \(\Omega\).  If the perturbation is nonaxisymmetric, the condition becomes \(Q<1.5-1.7\)  \citep{Durisen_review}.  Further to this condition, the cooling of the gas must be efficient enough to radiate away energy gained by compression during the contraction.  Use of the cooling time formalism allowed a quantitative statement of this second condition: \(t_{cool} \leq  3\Omega^{-1}\) \citep{Gammie,Ken_1,Mejia_2}. \\

\noindent More recent efforts have used more sophisticated approximations to capture the equation of state (EoS) of the gas, and model realistic radiative cooling and radiation transport, using both SPH \citep{WB_2,Mayer} and grid codes \citep{Mejia_4, BDNL}, but these are becoming increasingly complex, with some methods requiring mapping of the photosphere (which is often of non-trivial geometric shape), and extra conditions to be applied there (matching atmospheres as in \citet{Mejia_4}, or specifying cooling at the photosphere as in \citet{Mayer}).  Also, identifying the photosphere often requires extra free parameters, the changing of which will affect the final results.  The latest radiative transfer approximations (such as \citet{BDNL}, which solves the full radiative transfer equation explicitly in the vertical direction) are now attempting to remove this parametrisation. \\

\noindent This paper presents a new radiative transfer approximation, which relies on two separate methods working in tandem.  The first method is the polytropic approximation devised by \citet{Stam_2007}, which models the cooling of particles over a range of optical depths (\(0 < \tau \lesssim  10^{11}\)).  Its formulation ensures that cooling will be at its most efficient where the optical depth is around unity, in accordance with the definition of the photosphere.  This avoids the necessity of explicitly computing the location of the photosphere, or imposing boundary conditions upon it.  The second method is the flux-limited diffusion approximation, used by many authors \citep{boden_lambda,CM_diffuse,WB_1,WB_2,BDNL,Mayer,Mejia_4,Boss_2008} to simulate radiation transport in optically thick regimes.  Although presented as an algorithm for particle-based codes, the hybrid algorithm can be applied to grid-based codes also.  This hybrid method captures all the physics of frequency-averaged radiative transfer, without relying on parametrisation.\\

\noindent The paper is organised as follows. In section \ref{sec:Method} the constituents of the new hybrid algorithm are presented; it is then shown how these methods are combined to create the new algorithm.  In section \ref{sec:Tests}, the results from the following test scenarios are given: the evolution of a \(0.07\,M_{\odot}\) protoplanetary disc used as an example by \citet{Mejia_1}, \citet{Mejia_2}, \citet{Mejia_3}, and \citet{Mejia_4}; the collapse of a \(1M_{\odot}\) molecular cloud \citep{Masunaga_1}; and the thermal relaxation of a static sphere with seeded temperature fluctuations \citep{Spiegel,Masu_98}.  Finally, in section \ref{sec:Conclusions}, the method is summarised, and some indications of future work are given.

\section{Method }\label{sec:Method}

	\subsection{Polytropic Cooling}\label{sec:cooling}

\noindent The polytropic approximation uses an SPH particle's density \(\rho_i\), temperature \(T_i\), and gravitational potential \(\psi_i\) to estimate a mean optical depth for the particle \citep{Stam_2007}.  The approximation is achieved as follows.  Assume the particle is embedded in a spherically symmetric polytropic ``pseudocloud'' (which need not be in hydrostatic equilibrium).  The properties of the cloud are calculable analytically (using the Lane Emden equation), given the particle's (dimensionless) radius \(\xi\) from the centre: \(R=\xi R_0\).  Therefore, by appropriate selection of the central values of density and temperature \(\rho_c\), \(T_c\), and the scale-length \(R_0\), the particle's own values can be recovered:

\begin{equation} \rho_i = \rho_c\theta^n(\xi) 			\end{equation}
\begin{equation} T_i = T_c\theta(\xi) 	      			\end{equation}
\begin{equation}\psi_i = -4\pi G \rho_c R_{0}^{2}\phi(\xi).	\end{equation}

\noindent Here \(\theta\) is the solution to the Lane-Emden equation for a polytrope of index \(n\), and

\begin{equation} \phi(\xi) = -\xi_B \frac{d\theta}{d\xi}(\xi_B) + \theta(\xi) \end{equation}

\noindent (where \(\xi_B\) is the boundary of the polytrope) and \(R_0\) satisfies

\begin{equation} R_0 = \left[\frac{-\psi_i\theta^n(\xi)}{4\pi G \rho_i \phi(\xi)}\right]^{1/2}. \end{equation}

\noindent This provides the tools to calculate a column density from any given (dimensionless) radius to the boundary of the cloud:

\begin{equation} \Sigma_i(\xi) = \int^{\xi'=\xi_B}_{\xi'=\xi} \rho_c \theta^n(\xi')R_0 \,\mathrm{d}\xi' \end{equation}
\begin{equation} \Sigma_i(\xi) = \left[\frac{-\psi_i \rho_i}{4\pi G \phi(\xi) \theta^n(\xi)}\right]^{1/2} \int^{\xi'=\xi_B}_{\xi'=\xi} \theta^n(\xi') \,\mathrm{d}\xi' .\end{equation}

\noindent However, it is assumed that the value of \(\xi\) for the particle is unknown.  Instead, a value for the column density is arrived at by performing a \emph{mass weighted average} over all possible values of \(\xi\) out to the polytrope's boundary:

\begin{equation}\overline{\Sigma}_i =  \left[-\xi^2_B\frac{\mathrm{d}\theta}{\mathrm{d}\xi}(\xi_B)\right]^{-1} \int^{\xi'=\xi_B}_{\xi'=0} 
\Sigma_i(\xi') \,\theta^n(\xi')\xi^2 \,\mathrm{d}\xi' .\end{equation}

\noindent The total (dimensionless) mass of the polytrope is \(\left[-\xi^2_B\frac{\mathrm{d}\theta}{\mathrm{d}\xi}(\xi_B)\right]\), and \(\theta^n(\xi)\xi^2\mathrm{d}\xi\) is the dimensionless mass element between \(\left[\xi, \xi +\mathrm{d}\xi\right]\).  In real terms, \(\overline{\Sigma}_i\) becomes a simple algebraic quantity

\begin{equation}\overline{\Sigma}_i = \zeta_n \left[\frac{-\psi_i \rho_i}{4\pi G} \right]^{1/2} \end{equation}

\noindent with the integral folded into the constant \(\zeta_n\), which is dependent only on the polytropic index \(n\):

\begin{displaymath} \zeta_n=\left[-\xi^2_B\frac{\mathrm{d}\theta}{\mathrm{d}\xi}(\xi_B)\right]^{-1} \times\end{displaymath}
\begin{equation}\int^{\xi=\xi_B}_{\xi=0}\!\!\int^{\xi'=\xi_B}_{\xi'=\xi}\theta^n(\xi')\mathrm{d}\xi'\left[\frac{\theta^n(\xi)}{\phi(\xi)}\right]^{1/2}\!\!\xi^2 \mathrm{d}\xi .\end{equation}

\noindent \citet{Stam_2007} show that this constant is insensitive to the value of \(n\).  They select \(n=2\) for their work, as this would give a polytropic exponent of \(3/2\), in keeping with polytropic exponents of protostars in quasistatic equilibrium.  When using the polytropic formalism alone, the results of this paper will assume \(n=2\), (and hence \(\zeta_2 = 0.368\)) except where otherwise stated.  The simple expression for the column density illustrates its ability to capture the effects of the local environment (through the presence of \(\rho\)) and the effects of the system's global geometry (through the gravitational potential \(\psi\)). \\

\noindent In the same vein, a mass weighted optical depth can be calculated.  The optical depth from any radius to the edge of the pseudocloud is

\begin{equation} \tau_i(\xi) = \int^{\xi'=\xi_B}_{\xi'=\xi} \kappa_i\left(\rho_c \theta^n(\xi'),T_c \theta(\xi') \right) \rho_c \theta^n(\xi')R_0 \,\mathrm{d}\xi'	\end{equation}

\noindent Substituting for \(\rho_c\), \(T_c\) and \(R_0\) gives

\begin{displaymath} \tau_i(\xi) =\left[\frac{-\psi_i \rho_i \theta^n(\xi)}{4\pi G \phi(\xi)}\right]^{1/2} \times \end{displaymath}
\begin{equation} \int^{\xi'=\xi_B}_{\xi'=\xi} \kappa_i\left(\rho_i \left[\frac{\theta(\xi')}{\theta(\xi)}\right]^n ,T_i \left[\frac{\theta(\xi')}{\theta(\xi)}\right] \right)\left[\frac{\theta(\xi')}{\theta(\xi)}\right]^n \,\mathrm{d}\xi' .\end{equation}

\noindent Taking a mass weighted average then gives the rather messy

\begin{displaymath}\overline{\tau}_i = \left[-\xi^2_B\frac{\mathrm{d}\theta}{\mathrm{d}\xi}(\xi_B)\right]^{-1}\left[\frac{-\psi_i \rho_i}{4\pi G} \right]^{1/2} \int^{\xi=\xi_B}_{\xi=0} \int^{\xi'=\xi_B}_{\xi'=\xi} \end{displaymath}
\begin{equation}\kappa \left(\rho_i\left[\frac{\theta(\xi')}{\theta(\xi)}\right]^n, T_i\left[\frac{\theta(\xi')}{\theta(\xi)} \right]  \right) \theta^n(\xi')\mathrm{d}\xi' \left[\frac{\theta^n(\xi)}{\phi(\xi)}\right]^{1/2}\xi^2 \,\mathrm{d}\xi . \end{equation}	

\noindent This is a complicated function to calculate during a simulation. However, using the previous result for \(\overline{\Sigma}\), a mass weighted opacity can be defined,

\begin{equation} \overline{\kappa} = \frac{\overline{\tau}}{\overline{\Sigma}}, \end{equation}

\noindent and this can be evaluated in advance, and stored for later interpolation.  Hence, for a given \(\left( \rho,T \right)\):

\begin{displaymath}\overline{\kappa}(\rho,T) = \left[-\zeta_n\xi^2_B\frac{\mathrm{d}\theta}{\mathrm{d}\xi}(\xi_B)\right]^{-1} \int^{\xi=\xi_B}_{\xi=0} \int^{\xi'=\xi_B}_{\xi'=\xi} \end{displaymath}
\begin{equation}\kappa \left(\rho\left[\frac{\theta(\xi')}{\theta(\xi)}\right]^n, T\left[\frac{\theta(\xi')}{\theta(\xi)} \right]  \right) \theta^n(\xi')\mathrm{d}\xi' \left[\frac{\theta^n(\xi)}{\phi(\xi)}\right]^{1/2}\xi^2 \,\mathrm{d}\xi .\end{equation}	

\noindent The interpretation of this result is important: embedding the particle at some position in the polytrope ensures that the environment immediately surrounding the particle has a strong effect on its optical depth, and hence its emission.  This allows (for example) insulation of hot particles by cooler surroundings.  It is vital at this juncture to appreciate the meaning of this: \emph{the formalism is attempting to compensate for absorption of escaping radiation by modifying the net radiative losses} of the particles using the polytrope approximation. \\

\noindent If the net cooling term for SPH particle \emph{i} is \(\dot{u}_{i,cool}\), this then becomes

\begin{equation} \dot{u}_{i,cool} = \frac{4\sigma\left(T^4_0(\mathbf{r_i}) - T^4_i\right)}{\overline{\Sigma}^2_i \overline{\kappa}_i(\rho_i,T_i) + \kappa^{-1}_i(\rho_i,T_i)} .\end{equation}

\noindent The addition of \(T_0\) allows for external heating from a background radiation field (which can be configured to include irradiation from stellar objects).  Note that both the particle's opacity and mass weighted opacity are required.  The first term in the denominator becomes dominant in the optically thick case (where the particle's environment will absorb much of the cooling radiation it emits, reducing the energy loss), and the second term becomes dominant in the optically thin case (where the effects of the environment are less important, so the standard opacity is used).  Strictly speaking, the first term should be a mass weighted average of the Rosseland-mean opacity, and the second term should use the Planck-mean opacity, but in the case of this work the Rosseland-mean and Planck-mean opacities are taken to be equal.  An explanation of the two terms in the denominator can be found in \citet{Stam_2007}.\\

\noindent The construction of \(\dot{u}_{i,cool}\) allows the code to move smoothly from optically thin to optically thick regimes, and also identifies an optimum regime where the optical depth is of order unity, where the particle can emit radiation most efficiently (i.e. the photosphere). \\

\noindent The method is very efficient, having little impact on the total simulation time and performing very well in several tests of its ability \citep{Stam_2007}. Unfortunately, it does suffer from some key limitations:

\begin{enumerate}
\item Assuming a spherical pseudocloud will place restrictions on how well the code models different geometries: configurations that lack spherical symmetry will not be modelled as accurately as those that are spherically symmetric, although its general accuracy has been shown to be good \citep{Stam_2007}.
\item Although the formalism accounts for the surroundings of the particle when modelling its emission, it does not deal in detail with the exchange of heat energy between neighbouring fluid elements.  This makes it incapable of capturing accurately all the physics in the optically thick regime.
\end{enumerate}

	\subsection{Flux-limited Diffusion}\label{sec:diffusion}

\noindent The modelling of energy exchange is implemented using the flux-limited diffusion formalism used by \citet{Mayer}, which is in turn based on work in conduction modelling by \citet{CM_diffuse} and the flux limiter used by \citet{boden_lambda}.  In the diffusion approximation, the rate of energy change for particle \(i\) is

\begin{equation} \dot{u}_{i,diff} = \sum_b \frac{4m_b}{\rho_i \rho_b} \frac{k_ik_b}{k_i + k_b} (T_i - T_b) \frac{\mathbf{r}_{ib}.\underline{\nabla} \mathbf{W}}{|r_{ib}|^2}. \end{equation}

\noindent The summation index \(b\) describes the nearest neighbours of the particle (which is tracked by SPH to evaluate density fields and other requisite variables); \(\mathbf{W}\) is the smoothing kernel, \(\mathbf{r}_{ib}\) is the separation vector between particles \(i\) and \(b\), and \(k_i\) describes the thermal conductivity of the particle.  The gradient of the kernel is everywhere negative, so if \(T_i > T_b\), the summand will be negative (i.e. energy will flow from particle \(i\) to particle \(b\), in accordance with the laws of thermodynamics).  If the system's energy budget is defined entirely by diffusion, the particles will exchange energy amongst themselves in order to reduce temperature gradients; the long term evolution of the system will be towards a single equilibrium temperature. This ``washing out'' of temperature gradients is of critical importance: when simulating protoplanetary discs, the temperature profile (both radially and vertically) can define the regions of the disc where possible fragmentation can occur, and hence the regions where giant planets may form \citep{Boss_science}.  Any process which affects these profiles will influence where these regions are located.  \\

\noindent It should be noted at this point that all energy changes due to these diffusion terms are \emph{pairwise}, i.e. any energy loss by one particle will be matched by gain in its counterpart.  This means that the total energy change over the entire system due to diffusion must be zero:

\begin{equation}  \sum_i \dot{u}_{i,diff} = 0 .\end{equation}

\noindent This is an important feature, which allows it to be used in the hybrid method, as will be shown later.  The thermal conductivity is

\begin{equation} k_i = \frac{16\sigma}{\rho_i \kappa_i}\lambda_i T^3_i \end{equation}

\noindent where \(\kappa_i\) is the opacity, \(\sigma\) is the Stefan-Boltzmann constant and \(\lambda_i\) is the flux limiter. \citet{boden_lambda} describe an expression for \(\lambda_i\) which is calculated from the local radiation field:

\begin{equation}\lambda_i(R_i) = \frac{2+R_i}{6+3R_i+R_i^2}. \end{equation}

\noindent Here \(R_i\) is a function of the radiation energy density at the particle's position, \(u_r(\mathbf{r_i})\):

\begin{equation} R_i = \frac{|\underline{\nabla} u_r(\mathbf{r_i})|}{u_r(\mathbf{r_i})\rho_i \kappa_i} .\end{equation}

\noindent Studying the expression for \(R_i\), there are two limiting cases:

\begin{itemize}
\item When the region is very optically thick, \(\rho\) and \(\kappa\) become large (and the radiation field becomes uniform), and hence \(R_i \rightarrow 0\).  In this limit, the flux limiter \(\lambda_i \rightarrow 1/3\), in accordance with the diffusion approximation.
\item In the very optically thin limit, \(R_i\) becomes very large, and \(\lambda_i \rightarrow 0\), ending energy transport by diffusion.
\end{itemize}

\noindent This approximation is valid in the optically thick regime (and to lower optical depths with the use of the flux limiter, which prevents energy exchange as the mean free path of the radiation becomes prohibitively large).  Limitations of this method are:

\begin{enumerate}
\item It does not model radiation well at very low optical depths (where energy exchange disappears)
\item It does not allow the system to lose energy (i.e. it does not model radiative cooling).  Instead, this cooling must be added using a prescription which assumes prior knowledge of the geometry of the system being studied, and invokes resolution dependent free parameters (e.g. \citealt{Mayer}).
\end{enumerate}

	\subsection{The Hybrid Method}

\noindent Comparing the limitations of the above two methods, it is clear that a union of these two procedures should be complementary: polytropic cooling handles the important energy loss from the system (which flux-limited diffusion cannot), and flux-limited diffusion handles the detailed exchange of heat between neighbouring fluid elements (which polytropic cooling cannot).  Indeed, polytropic cooling's inability to model the detailed exchange of heat between neighbouring fluid elements - and flux-limited diffusion's inability to model energy loss - allow the two methods to work together correctly, modelling all aspects of the system's energy budget without encroaching on each other.  The energy equation simply becomes

\begin{equation} \dot{u}_{i,total} = \dot{u}_{i,hydro} + \dot{u}_{i,cool} + \dot{u}_{i,diff} ,\end{equation}

\noindent where \(\dot{u}_{i,hydro}\) describes the energy change due to the hydrodynamics of the system, e.g. compressive \(P\mathrm{d}V\) heating.  The true advantage to using this hybrid method is in its simplicity:

\begin{itemize}
\item By construction, the hybrid method is fully three-dimensional, and capable of handling arbitrary particle geometries.
\item There is no requirement to grid the system.
\item The algorithm is continuous over a wide range of optical depths, so there are no requirements to match separate atmospheres at some boundary.
\item As no extra boundary conditions are required, there are no extra parameters to be specified, so the simulation's results are only dependent on the traditional SPH parameters (particle number, smoothing length etc).
\end{itemize}

\noindent However, the hybrid method still suffers from some disadvantages:

\begin{enumerate}
\item This method is still unable to model frequency dependent radiative transfer.
\item As with polytropic cooling, the hybrid method is better suited to modelling the cooling of spherical geometries.
\end{enumerate}

	\subsection{Updating Energy: A Semi-Implicit Scheme}

\noindent The use of an explicit scheme to update energy can result in very short time steps.  To avoid this, a modified version of the implicit scheme adopted by \citet{Stam_2007} is used.  This models each particle's approach to its equilibrium temperature \(T_{eq}\), which satisfies

\begin{equation} \dot{u}_{i,hydro} + \frac{4\sigma\left(T^4_0(\mathbf{r_i}) - T_{eq,i}^4\right)}{\overline{\Sigma}^2_i \overline{\kappa}_i(\rho_i,T_{eq,i}) + \kappa^{-1}_i(\rho_i,T_{eq,i})} + \dot{u}_{i,diff} = 0 .\end{equation}

\noindent From this, the equilibrium internal energy \(u_{eq,i}=u(\rho_i,T_{eq,i})\) can be calculated, and hence the thermalisation timescale:

\begin{equation} t_{therm} = \frac{u_{eq,i}-u_{i}}{\dot{u}_{i,total}}. \end{equation}

\noindent With knowledge of how quickly each particle can be thermalised, the particle's energy can be updated thus:

\begin{equation} u_i(t+\Delta t) = u_i(t)exp\left[\frac{-\Delta t}{t_{therm}}\right] + u_{eq,i}\left(1 - exp\left[\frac{-\Delta t}{t_{therm}}\right] \right) \end{equation}

\noindent For particles that will thermalise very quickly (\(t_{therm} \ll \Delta t\)), which would result in very short timesteps, this equation reduces to

\begin{equation} u_i(t+\Delta t) \approx u_{eq,i}\end{equation}

\noindent i.e., the particle rapidly reaches equilibrium.  If thermalisation happens on a long timescale (\(t_{therm} \gg \Delta t\)), then the equation becomes

\begin{equation} u_i(t+\Delta t) \approx u_i(t) + \left(u_{eq,i} - u_i(t)\right)\frac{\Delta t}{t_{therm}} \end{equation}

	\subsection{Properties of the Dust and Gas}

\begin{figure}
\begin{center}
\includegraphics[scale = 0.5]{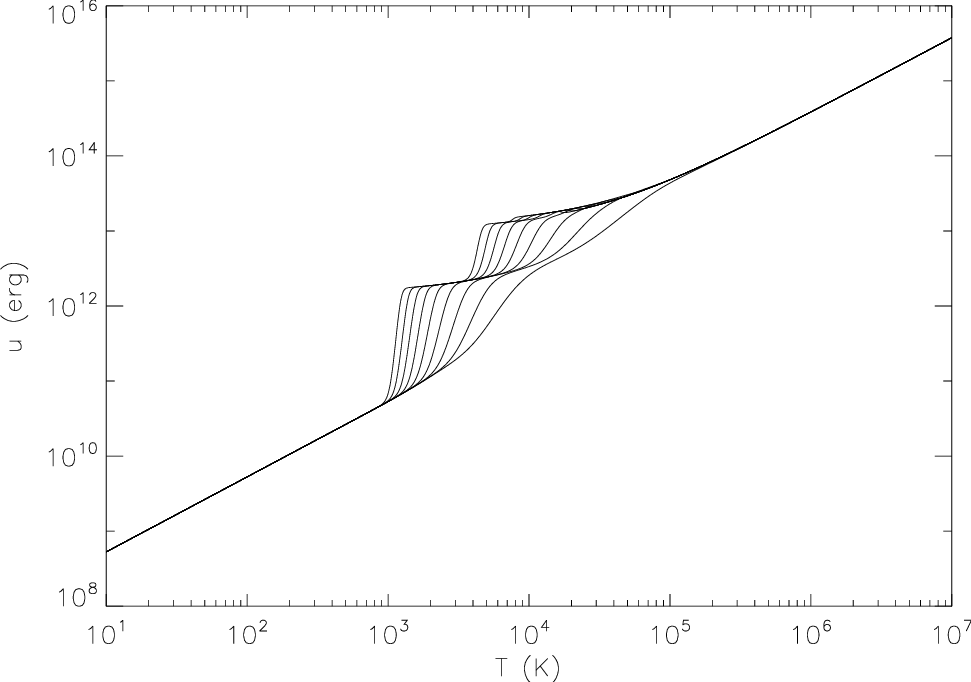}
\caption{Internal energy u as a function of T for various densities.  Curves are plotted for \(\rho = 10^{-18} \,g\, cm^{-3}\) (bottom curve) to \(\rho = 10^{-2} \,g\, cm^{-3}\) (top curve).\label{fig:u_plot}}
\end{center}
\end{figure}

\begin{figure}
\begin{center}
\includegraphics[scale = 0.5]{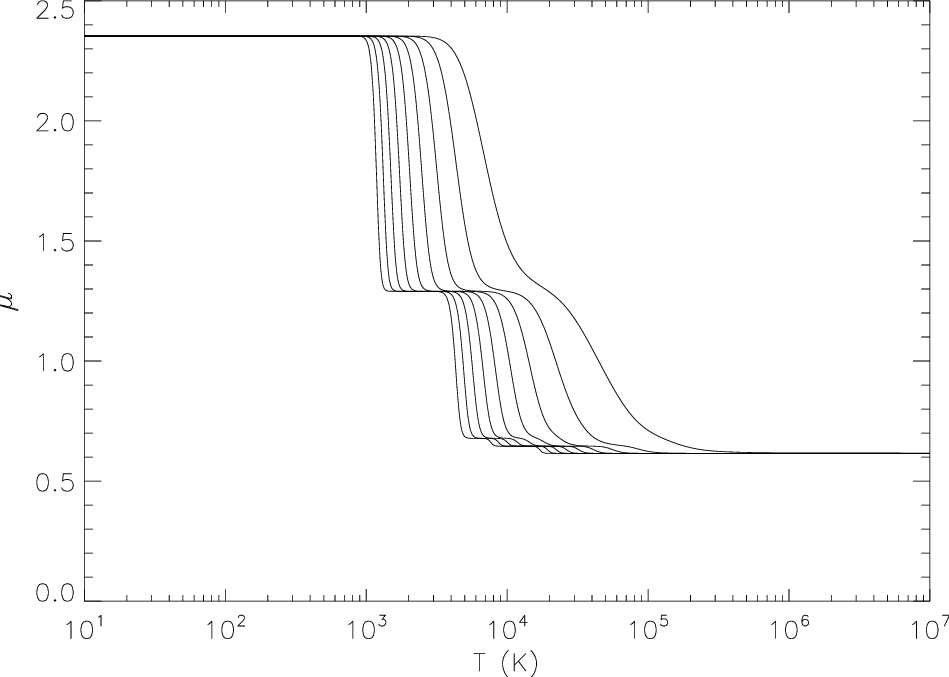}
\caption{Molecular weight as a function of T for various densities. Curves are plotted for \(\rho = 10^{-18} \,g\, cm^{-3}\) (bottom curve) to \(\rho = 10^{-2} \,g\, cm^{-3}\) (top curve).\label{fig:muplot}}
\end{center}
\end{figure}

\begin{figure}
\begin{center}
\includegraphics[scale = 0.5]{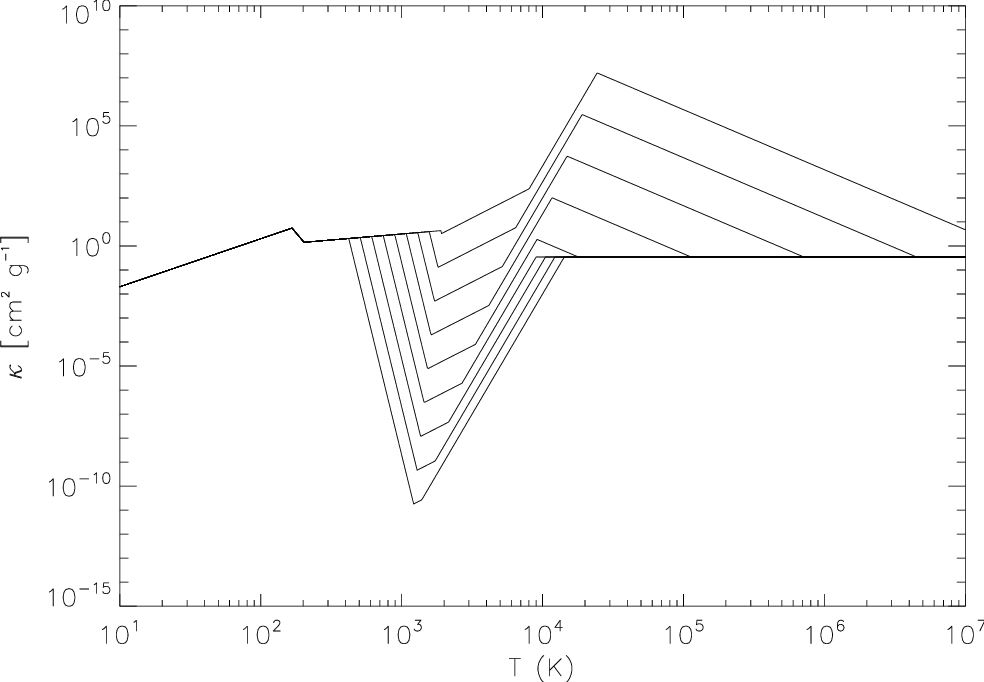}
\caption{Rosseland mean opacity as a function of temperature for a series of different densities. Curves are plotted for \(\rho = 10^{-18} \,g\, cm^{-3}\) (bottom curve) to \(\rho = 10^{-2} \,g\, cm^{-3}\) (top curve) \label{fig:kappa}.}
\end{center}
\end{figure}

\begin{figure}
\begin{center}
\includegraphics[scale = 0.5]{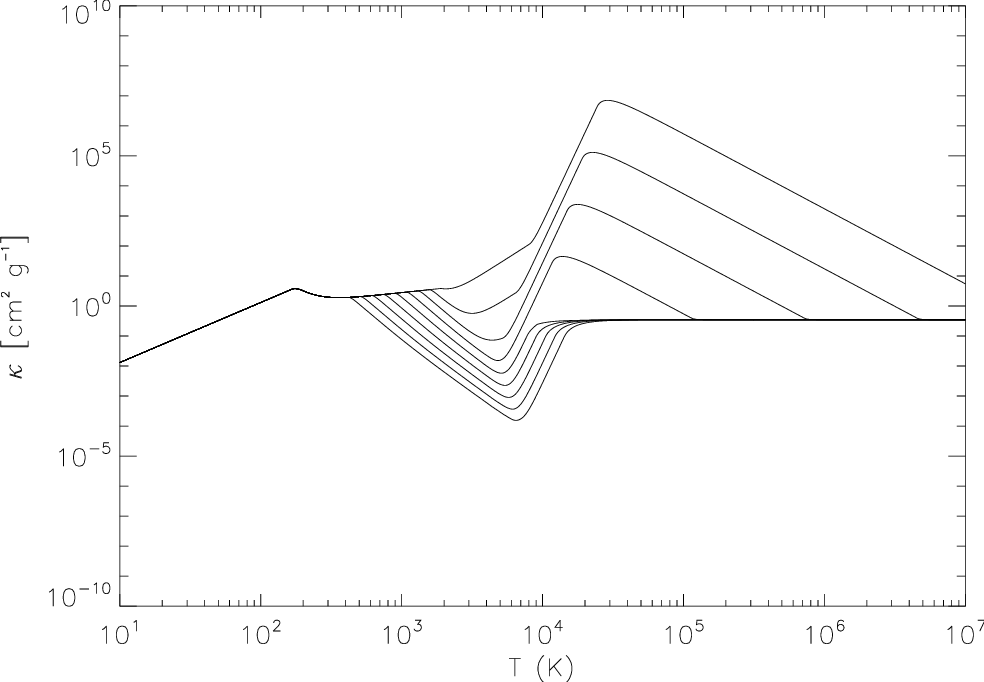}
\caption{Mass weighted opacity as a function of temperature for a series of different densities. Curves are plotted for \(\rho = 10^{-18} \,g\, cm^{-3}\) (bottom curve) to \(\rho = 10^{-2} \,g\, cm^{-3}\) (top curve) \label{fig:kappa_bar}.}
\end{center}
\end{figure}

\noindent Vital to any radiative transfer method is how the variables it uses (temperature, opacity, etc) are evaluated.  SPH evolves only the density and internal energy of the particles: hence some kind of prescription is required in order to obtain the requisite data.  Essentially, what is required is \(T(\rho,u)\), \(\kappa(\rho,u)\) etc.  In practice, it is more straightforward to evaluate \(u(\rho,T)\) (known as the equation of state) and \(\kappa(\rho,T)\) (the opacity law) and tabulate these values, which can then be interpolated to achieve the correct results.  The equation of state and the opacity law used in this work are similar to that of \citet{Stam_2007}, where a full description is available (see also \citealt{Black,Boley_hydrogen}).  This work assumes hydrogen and helium mass fractions of \(X = 0.7\), \(Y=0.3\), and a fixed ortho- to para-hydrogen ratio of \(3:1\).  The dependence of the various variables on temperature can be seen in Figures \ref{fig:u_plot}, \ref{fig:muplot}, \ref{fig:kappa} \& \ref{fig:kappa_bar}.  Figures \ref{fig:u_plot} \& \ref{fig:muplot} show the activation of the various energy states of hydrogen and helium gas as temperature increases: Figure \ref{fig:kappa} shows the opacity law as calculated according to the prescription of \citet{Bell_and_Lin}.  Figure \ref{fig:kappa_bar} shows the mass weighted opacity as discussed in \citet{Stam_2007}.  The opacity law captures many different opacity regimes of the gas, including ice and dust opacities, as well as molecular absorption, bound-free and free-free interactions and electron scattering.

\section{Tests }\label{sec:Tests}

\noindent The code used to perform these tests is based on the SPH code developed by \citet{Bate_code}.  It uses variable individual smoothing lengths \(h_i\) in order that the number of nearest neighbours for any particle is \(50\pm20\).  It uses individual particle timesteps to allow dense regions to be simulated with greater time resolution while preventing oversimulation of less dense regions.  A binary tree is employed to calculate neighbour lists and calculate gravity forces.  The standard artificial viscosity is also used.  All simulations are sufficiently populated to satisfy the Jeans resolution condition of \citet{Burkert_Jeans} for Jeans masses of \(30\,M_{\oplus}\) or less.  These conditions are sufficient for the cloud simulations performed.\\

\noindent For the disc simulations performed, the Toomre length becomes important in the regions that are unstable, and places stricter resolution conditions.  As the disc simulation is Keplerian, the following relation can be used between the Jeans length and the Toomre length \citep{Nelson_res}:

\begin{equation} \lambda_T = \sqrt{\frac{2Q}{f}} \lambda_J, \end{equation}

\noindent where \(f\sim 1\) represents the conversion factor between surface and volume densities.  As the disc is marginally unstable (\(Q\sim 1\)), the Toomre length can be simply calculated.  Converting this (assuming a homogeneous sphere) into a Toomre Mass, it is calculated that the disc simulation can resolve Toomre masses of \(\sim 85\,M_{\oplus}\) or more.

	\subsection{The Evolution of a Protoplanetary Disc}

\noindent As a means of comparison with previous results, the conditions used for this test are those proposed by Mej\'{i}a et al, and used in a series of papers describing radiative transfer in protoplanetary discs \citep{Mejia_1, Mejia_2, Mejia_3, Mejia_4}.  The model is a \(0.07\,M_{\odot}\) Keplerian disc which extends to 40 AU, orbiting a star of \(0.5\,M_{\odot}\).  Initially, the surface density profile is \(\Sigma \sim r^{-1/2}\), with a temperature profile of \(T \sim r^{-1}\).  The disc is modelled using \(2.5 \times 10^5\) SPH particles, with one sink particle representing the star.  The disc is immersed in a radiation field of \(T_0(\mathbf{r}) = 3K\) ; the effects of disc irradiation by the central star are not included. \\

\begin{figure*}
\begin{center}$
\begin{array}{cc}
\includegraphics[scale = 0.39]{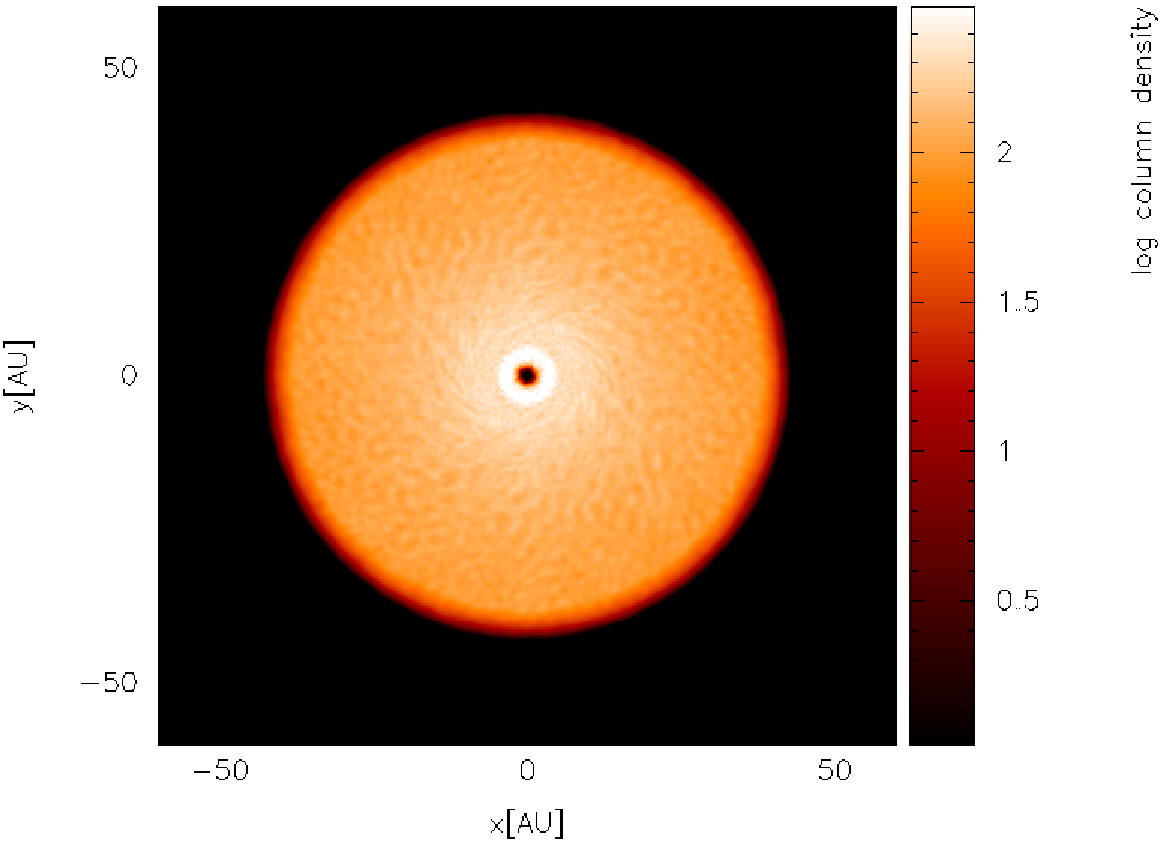} &
\includegraphics[scale = 0.39]{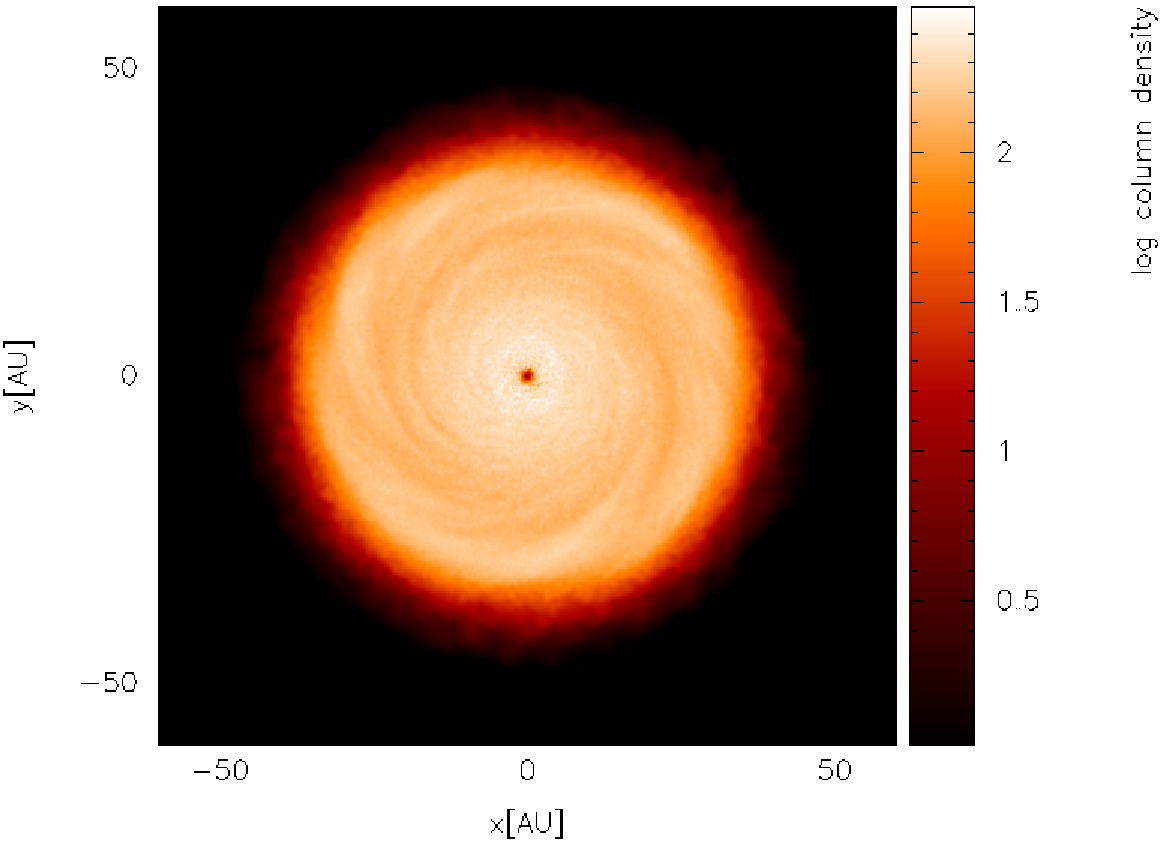} \\
\includegraphics[scale = 0.39]{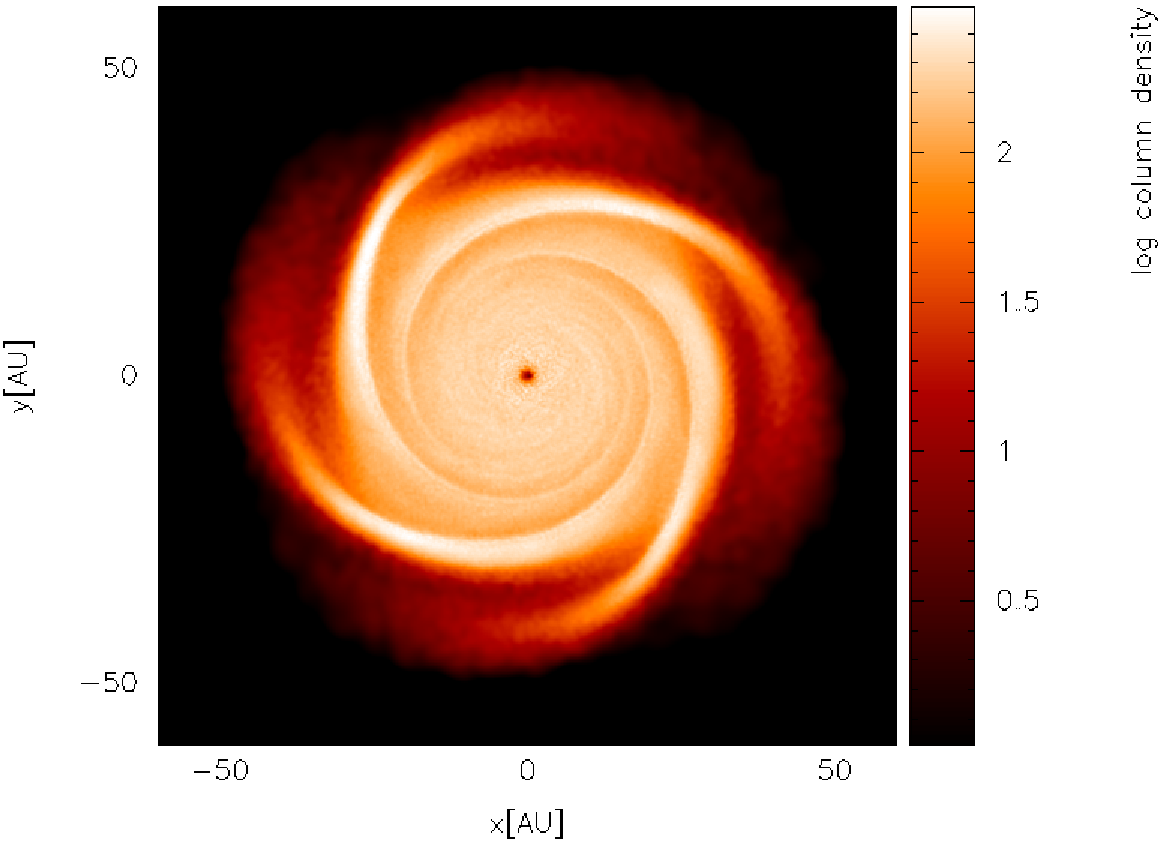} &
\includegraphics[scale = 0.39]{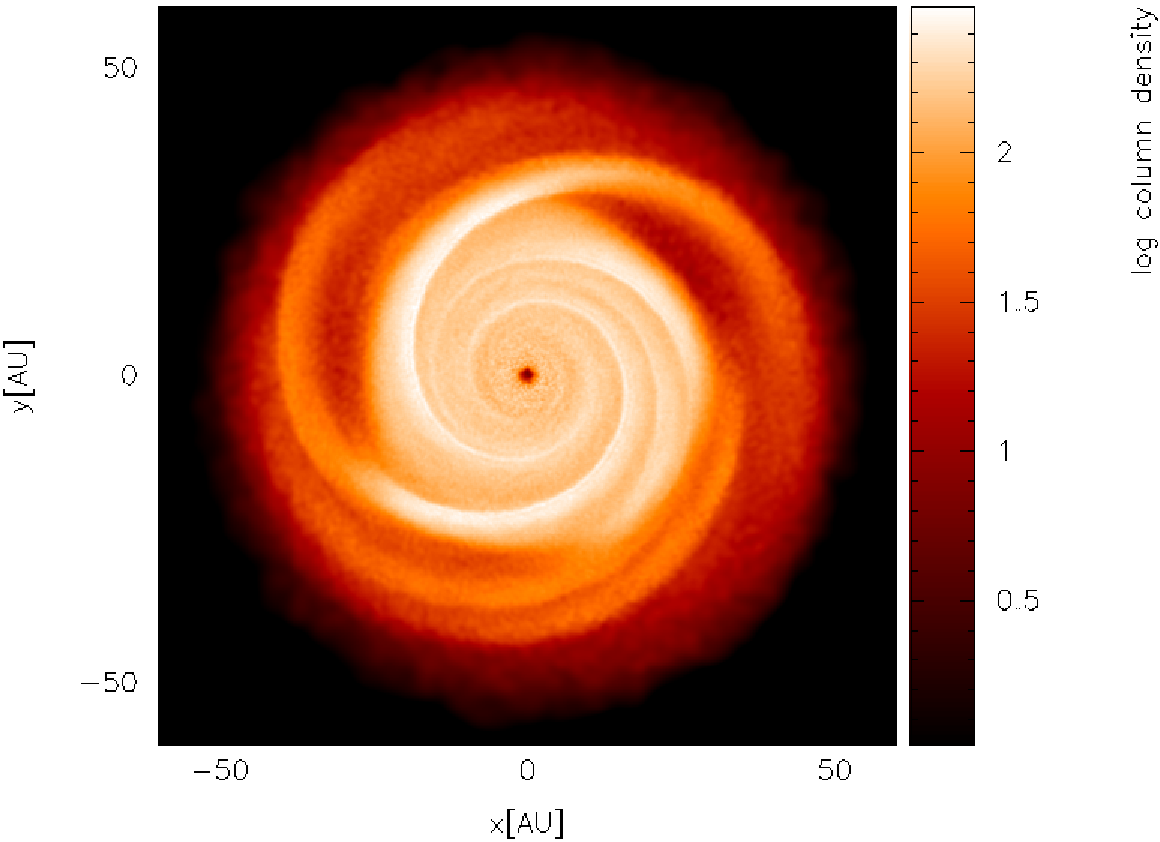} 
\end{array}$
\caption{Evolution of the Boley disc at various times under the hybrid method \label{fig:Mejia_discs}.  The images are taken at the following times: 9.72 years (top left), 506 years (top right), 992 years (bottom left), 1906 years (bottom right).}
\end{center}
\end{figure*}	 	

\begin{figure*}
\begin{center}$
\begin{array}{cc}
\includegraphics[scale = 0.39]{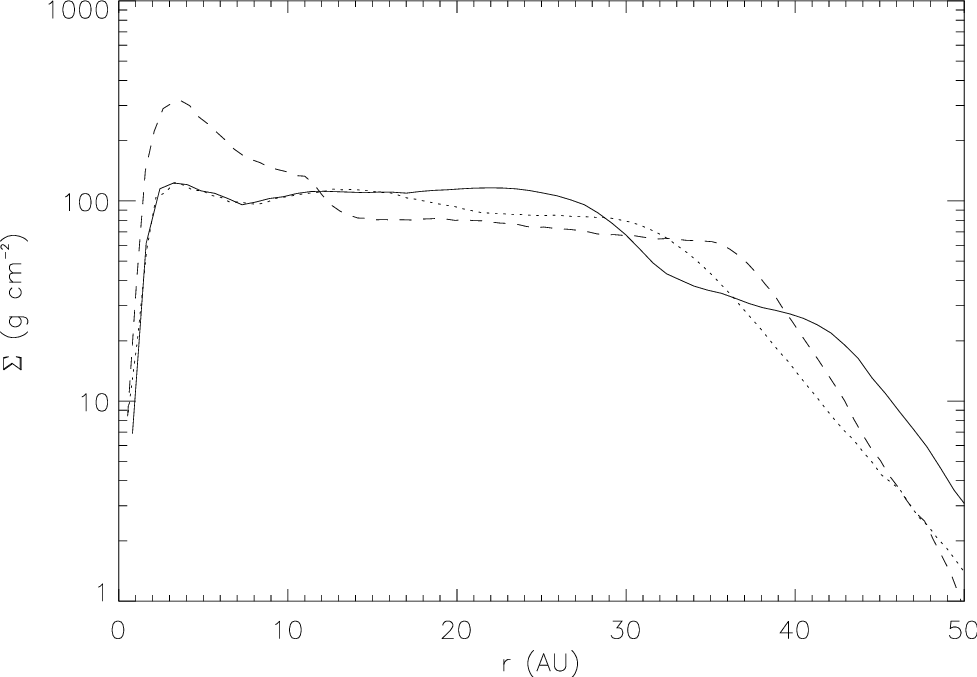} &
\includegraphics[scale = 0.39]{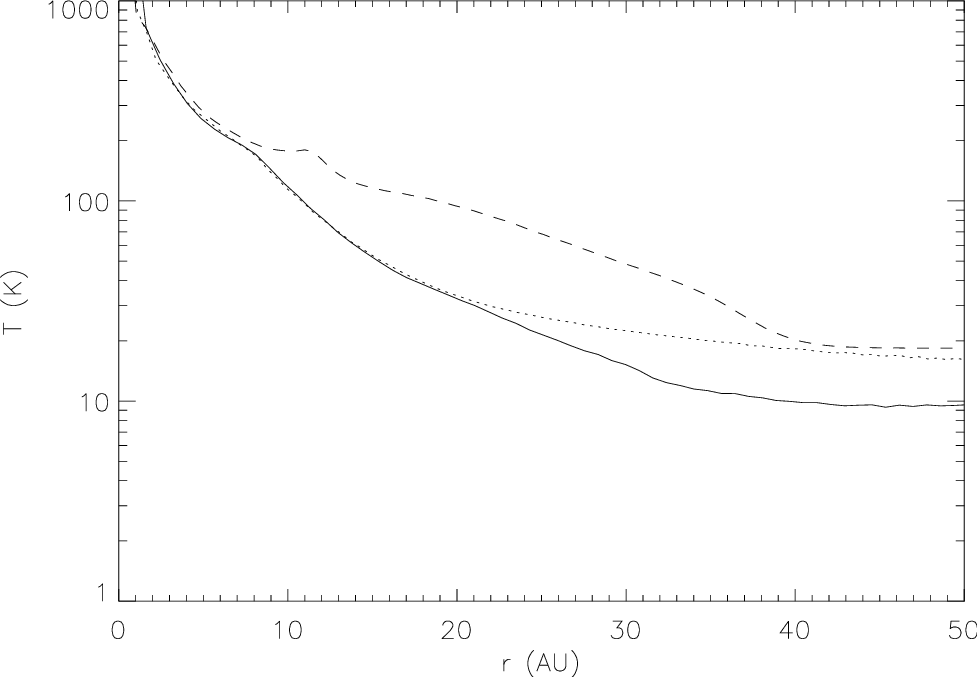} \\
\includegraphics[scale = 0.39]{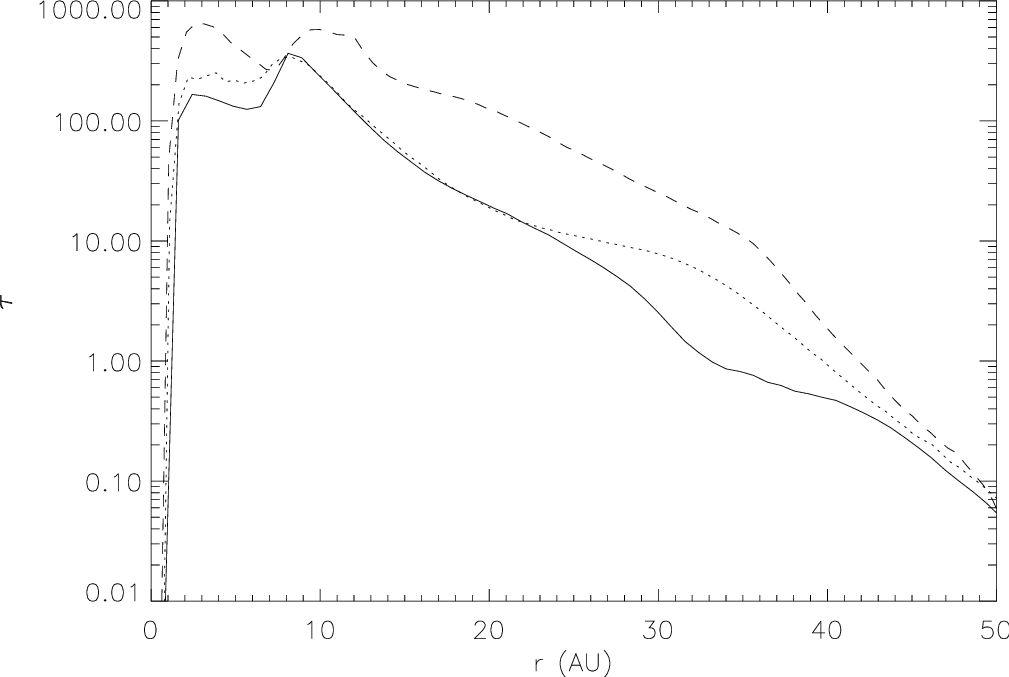} &
\includegraphics[scale = 0.39]{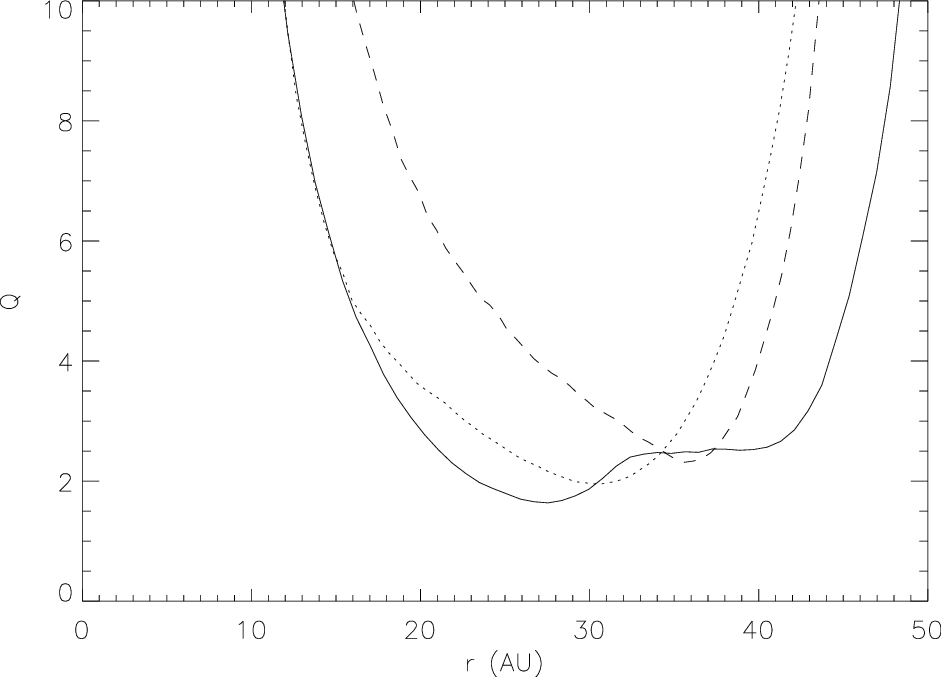} 
\end{array}$
\caption{Azimuthally averaged radial profiles of the Boley disc at \(t = 1906\) years: the solid lines are the results obtained using the hybrid method, the dotted lines are the results obtained using the polytropic cooling approximation alone, and the dashed lines are the disc at $t = 9.72$ years using the hybrid method.  The top left panel shows the surface density of the disc; the top right panel shows the midplane temperature of the disc; the bottom left panel shows the optical depth from the midplane to the disc surface; the bottom right panel shows the Toomre instability parameter. \label{fig:Mejia_profiles}}
\end{center}
\end{figure*}	 

\begin{figure*}
\begin{center}$
\begin{array}{cc}
\includegraphics[scale=0.4]{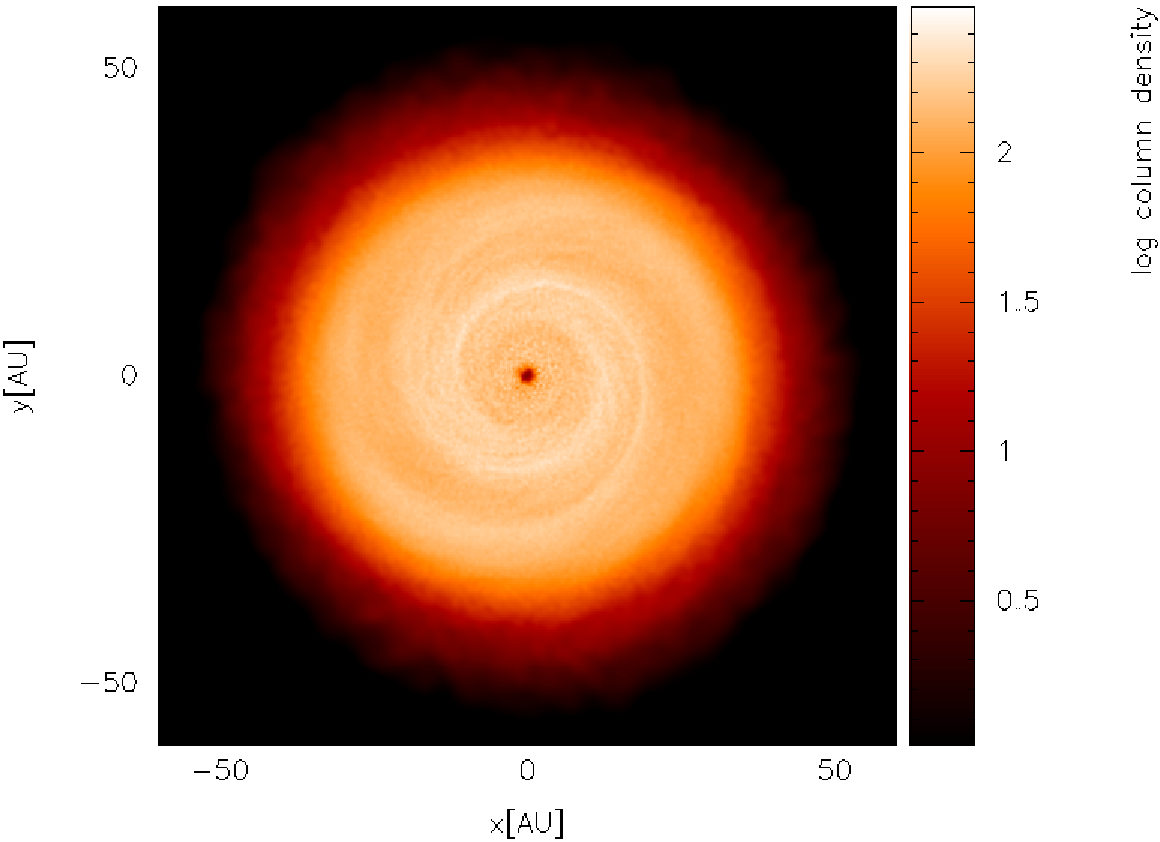} &
\includegraphics[scale=0.4]{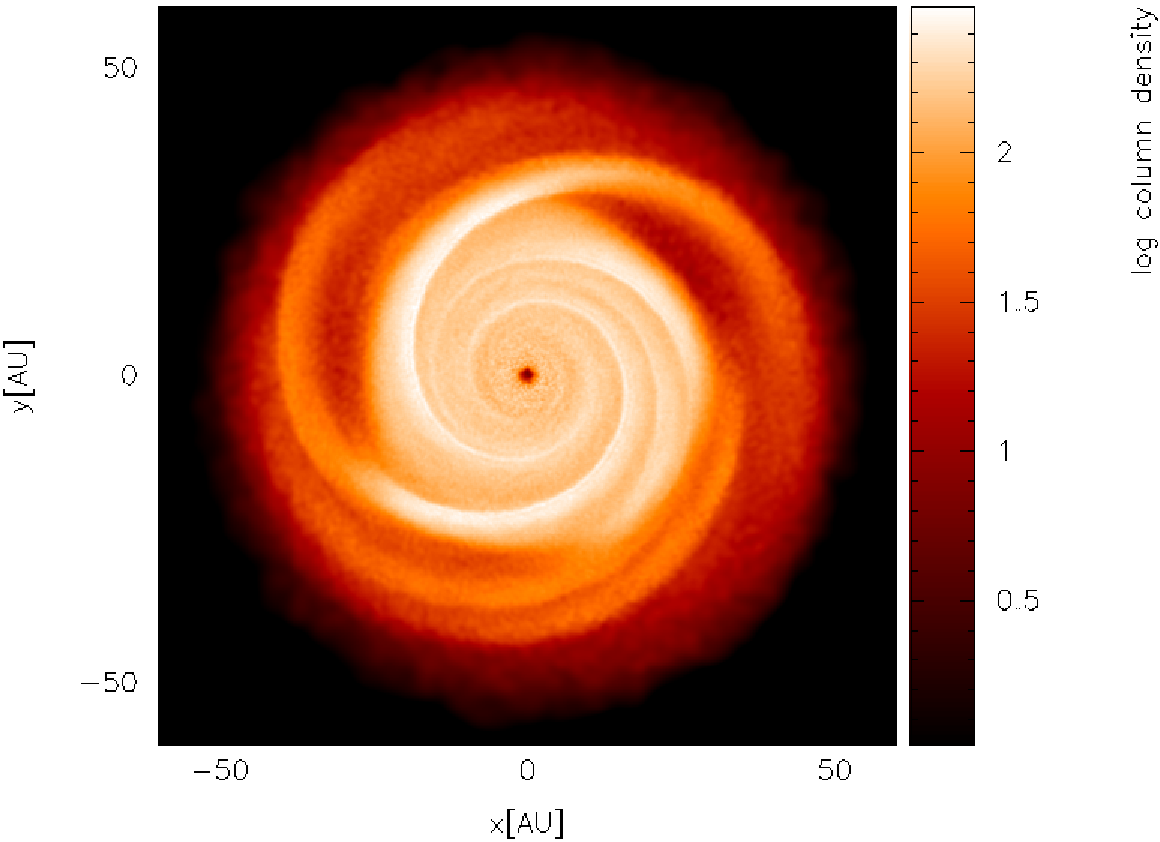} \\
\end{array}$
\caption{Comparing the hybrid method and polytropic cooling for the Mej\'{i}a disc at $t = 1906$ years.  The left panel shows the evolved disc under polytropic cooling alone; the right panel shows the evolved disc under the hybrid method. \label{fig:twodisc}}
\end{center}
\end{figure*}

\noindent The properties of the evolved disc using both the hybrid method and the polytropic cooling approximation alone are shown in \ref{fig:Mejia_profiles}.  For both methods, several key phases are identified: the initial settling phase, during which the disc adjusts its outer radius by axisymmetric evolution, and ring formation and contraction occurs; the ``burst'' phase where nonaxisymmetric instabilities, in the form of spiral waves, begin to grow; and the later asymptotic phase, where the disc's radial extent is more firmly established, and the gravitational instability is regulated (cf. \citealt{Mejia_3}).  As the evolution of this asymptotic phase continues, the low-m modes begin to dominate.  In terms of timescale, the settling phase lasts until  \(t \sim 500\) years, the burst phase until \(t \sim 1200\) years, where the asymptotic phase then begins, resulting in a quasi-equilibrium state.  \\

\noindent Comparing the hybrid method against the results of using the polytropic cooling approximation alone, there are significant differences: the hybrid method transports more mass radially outward, which can be seen in the surface density of the disc (Figure \ref{fig:Mejia_profiles}, top left panel).  This has several important consequences: it allows the optical depth to be reduced in the region \(r\sim 20 - 40\) AU, which allows an increase in radiative cooling; this in turn allows the outer disc to be cooler, and for the outer regions of the disc (\(r > 20\) AU) to become less stable (as can be seen in the other panels of Figure \ref{fig:Mejia_profiles}).  Snapshots of the disc under both methods can be seen in (Figure \ref{fig:twodisc}).  Note the stronger spiral structure in the disc under the hybrid method, with instabilities extending to larger radii.  All these differences are critical if the formation of giant planets by gravitational instability is to be effectively tested by simulation.  \\

\noindent Comparing the hybrid method to the results of \citet{Mejia_3,BDNL}, the two are qualitatively consistent.  Each has a burst phase, and an asymptotic phase; each has a two-component surface density profile (approximately flat at lower radii, with a cut off at larger radii).  There are also some quantitative consistencies: the optical depth from the midplane to the surface in the hybrid method reaches unity at \(R\sim 27 \) AU, which is coincident with the region of the disc that is most unstable (i.e. the Toomre Q parameter is at a global minimum) - which is in keeping with the work of Boley et al.  It can also be seen (by comparing the surface density profiles of the hybrid method and polytropic cooling) that there appears to be a surplus of matter within \(R\sim 20 -  27\) AU, and a slight deficit at \(R\sim 27-40\) AU, indicating that \(R\sim 27\) AU may be the location where mass transport switches from inward to outward, again consistent with the results of Boley et al.  However, there are some important differences to be considered.  The burst phase of the hybrid method is noticeably weaker, and the disc undergoes less radial spreading.  This also means that in the asymptotic phase, the unstable region is much narrower in radius.  The first component of the surface density profile also appears to be flatter at lower radii for the hybrid method.  \\

\noindent It should be noted at this point that there are mitigating factors at work: the equation of state and opacity law used in this work is different from that of Boley et al; also, they fix the star at the centre of their grid: the star used in these results is allowed to move.  The differences in the EoS and the opacity law will have a stronger effect in the hotter inner regions of the disc, perhaps explaining the differences in surface density profile, and the lack of radial spreading.  It should also be noted that the inner disc stays somewhat hotter than expected (for both polytropic cooling alone, and for the hybrid method).  This may be due to SPH viscosity: as the distance to the centre decreases, the magnitude of the SPH viscosity increases, and may become significant (relative to the effective gravitational viscosity).  This possibility will be investigated in more detail at a later date. \\

\noindent Although exciting spiral waves in all three cases (polytropic cooling alone, the hybrid method and the work of Boley et al), the instability in the disc does not lead to fragmentation.  Also, the disc is only Toomre unstable at larger radii, which does not bode well for \emph{in situ} formation of Jovian objects at \(R \leq 20\) AU (at least in these conditions).

	\subsection{The Collapse of a \(1\,M_{\odot}\) Cloud}

\noindent The collapse of a non-rotating molecular cloud was then simulated.  The spherical, uniform density cloud contains \(1\,M_{\odot}\) of material (populated by \(5 \times 10^5\) SPH particles), and has a radius of \(10^4\) AU (which gives a density of \(\rho_0 = 1.41 \times 10^{-19} \, \mathrm{g\, cm^{-3}}\)), and is immersed in a background radiation field of \(T_0(\mathbf{r}) = 5K\).  These conditions were initially investigated by \citet{Masunaga_1} by solving the full radiative transfer in 3D (with the hydrodynamics solved in 1D), and were revisited by \citet{Stam_2007}.  These conditions therefore represent not only a solid test of the code's ability to match Masunaga \& Inutsuka's data, but also allow us to compare with the results of Stamatellos et al to identify the effects of adding flux-limited diffusion. \\

\begin{figure*}
\begin{center}$
\begin{array}{cc}
\includegraphics[scale = 0.5]{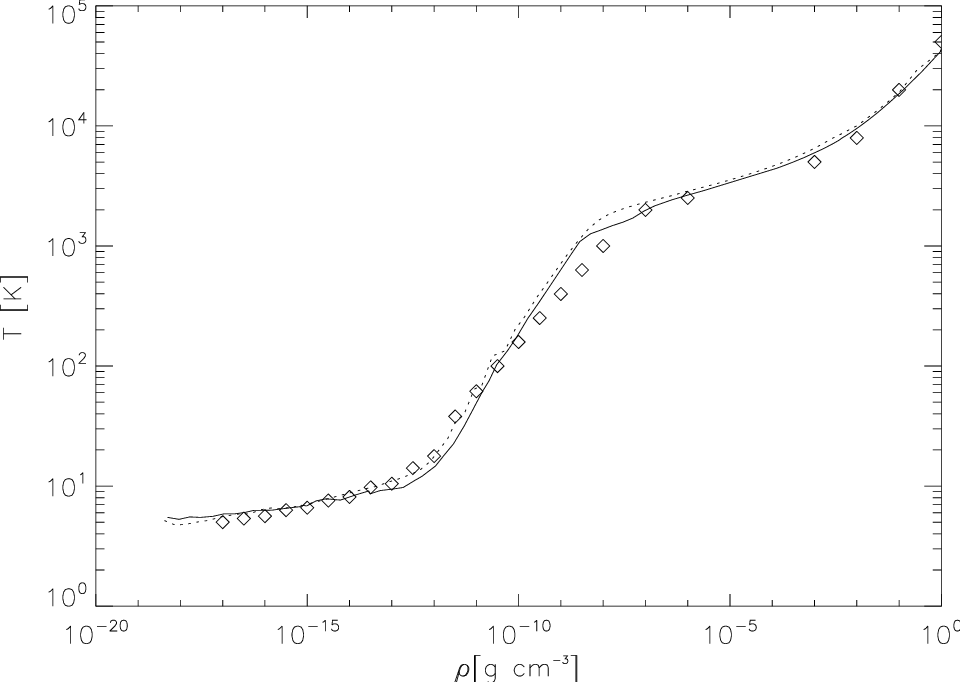} &
\includegraphics[scale = 0.5]{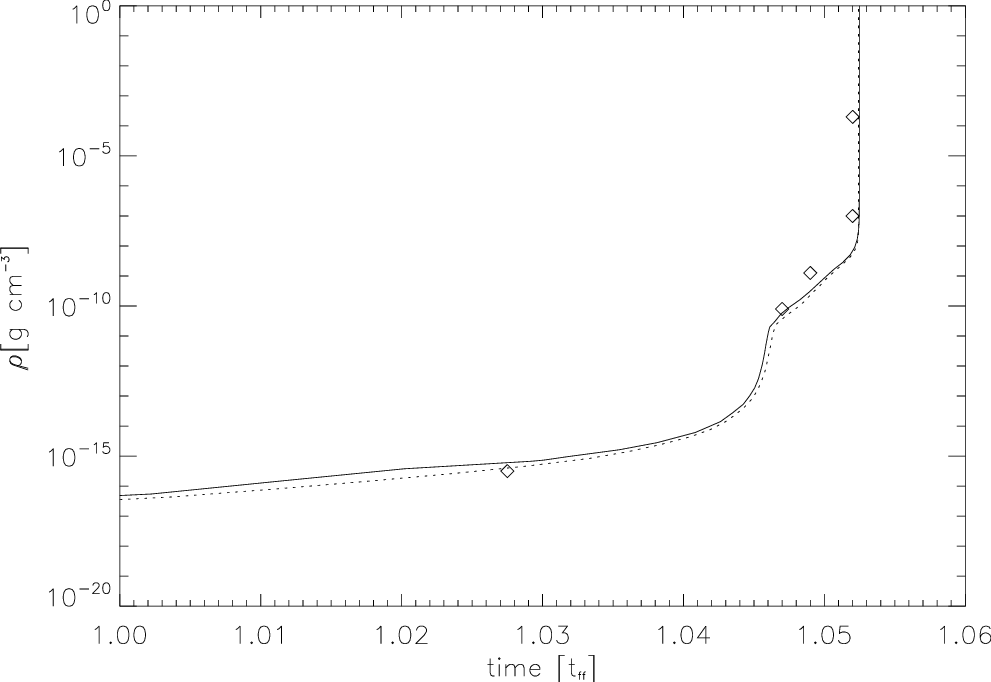} \\
\end{array}$
\caption{Evolution of the central density of the Masunaga Cloud - the left panel shows the evolution of central temperature with increasing central density,  the right panel shows the time evolution of the central density.  The solid lines represent the hybrid method, the dotted lines represent polytropic cooling only, and the diamonds represent the data of \citet{Masunaga_1}.   \label{fig:Masunaga_rhovsT}}
\end{center}
\end{figure*}

\noindent In the initial phase, the collapse is isothermal: the temperature remains at approximately 5 K through seven orders of magnitude in density (see Figure \ref{fig:Masunaga_rhovsT}, left panel), until the central density reaches \(\rho \sim 10^{-12} \, \mathrm{g\, cm^{-3}}\). The cloud then becomes optically thick, and the temperature starts to rise.   As the temperature reaches \(T\sim 100\) K, the rotational degrees of freedom of molecular hydrogen are activated, slowing the temperature increase slightly (this can be seen in the small bump in the left panel of Figure \ref{fig:Masunaga_rhovsT}).  The increased heating in the centre eventually decelerates the contraction at around \(\rho = 10^{-9} \, \mathrm{g\, cm^{-3}}\), and the first core is formed.  The contraction and heating of this core proceed until the central temperature is around \(T \sim 2000\) K.  The H$_2$ present begins to dissociate, using some of the available compressive energy due to the contraction.  This allows a second collapse, which can continue until most of the H$_2$ is dissociated.  After this the contraction decelerates again at around \(\rho = 10^{-3} \, \mathrm{g\, cm^{-3}}\), and the second core forms. \\

\noindent The dotted line in the left hand panel of Figure \ref{fig:Masunaga_rhovsT} shows the evolution of the Masunaga Cloud using polytropic cooling alone.  Both methods approximate the data of \citet{Masunaga_1} well (diamonds in Figure \ref{fig:Masunaga_rhovsT}).  However, there are two key differences: the hybrid run stays isothermal to slightly higher densities (where the extra loss of energy along temperature gradients due to diffusion keeps the cooling efficient enough to allow this), and the slight bump at \(\rho_0 \sim 10^{-9} \, \mathrm{g\, cm^{-3}}\) (again diffusion allowing the centre to cool more efficiently).  This demonstrates that the polytropic cooling method alone provides a good approximation of the energy exchange between neighbouring particles by correctly modelling the net radiative losses; the addition of flux-limited diffusion constitutes only a small additional exchange of energy. \\

\noindent The time evolution of the cloud (Figure \ref{fig:Masunaga_rhovsT}, right panel) follows closely the evolution described by \citet{Stam_2007} and \citet{Masunaga_1}.  As with Stamatellos et al, there are discrepancies with Masunaga and Inutsuka's data due to the use of different opacities, and slight variations in initial conditions.  By synchronising the simulations at a central density of \(\rho = 4.34 \times 10^{-13} \, \mathrm{g\, cm^{-3}}\) \citep{Stam_2007}, good agreement is obtained.     

\subsection{The Spiegel Test}

\noindent As a final test, the thermal relaxation of a static, spherical cloud with a well-defined temperature perturbation allows comparison of the hybrid method with analytic results.  The cloud is uniform in density, with \(\rho = 10^{-19}\mathrm{g \, cm^{-3}}\), and a radius of \(R=10^4\) AU.  The equilibrium temperature is taken to be \(T_0 = 10\) K, and an initial temperature perturbation which satisfies

\begin{equation} T(r) = T_0 + \Delta T_0 \frac{\sin{kr}}{kr} \end{equation}

\noindent where \(\Delta T_0 = 0.15\) K is the amplitude, and \(k = \frac{\pi}{2500\, \mathrm{AU}}\) is the characteristic wavenumber \citep{Spiegel,Masu_98}.  At a later time \(t\), this perturbation evolves according to \citep{Masu_98}

\begin{equation} T(r,t) = T_0 + \Delta T_0 \frac{\sin{kr}}{kr} e^{-\omega(k)t}.  \label{eq:spT}\end{equation}

\noindent In Equation (\ref{eq:spT}), 

\begin{equation} \omega(k) = \gamma \left[1- \frac{\kappa_0}{k} cot^{-1}\left(\frac{\kappa_0}{k}\right)\right] \end{equation}

\noindent and

\begin{equation} \gamma = \frac{16\sigma \kappa_0 T_0^3}{\rho c_v}.\label{eq:spgamma} \end{equation}

\noindent Here \(\kappa_0\) is the opacity at equilibrium and \(c_v\) is the heat capacity of the material.  This test was also performed by \citet{Stam_2007}, and hence provides an extra means of comparing polytropic cooling and the hybrid method.  \\

\begin{figure*}
\begin{center}$
\begin{array}{cc}
\includegraphics[scale=0.5]{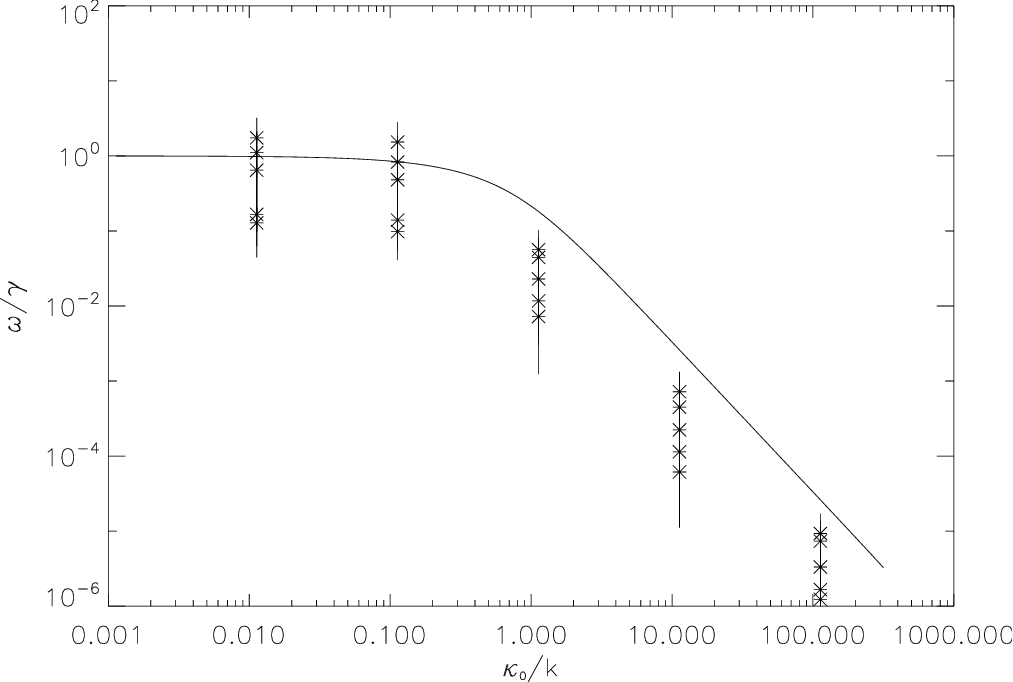} &
\includegraphics[scale=0.5]{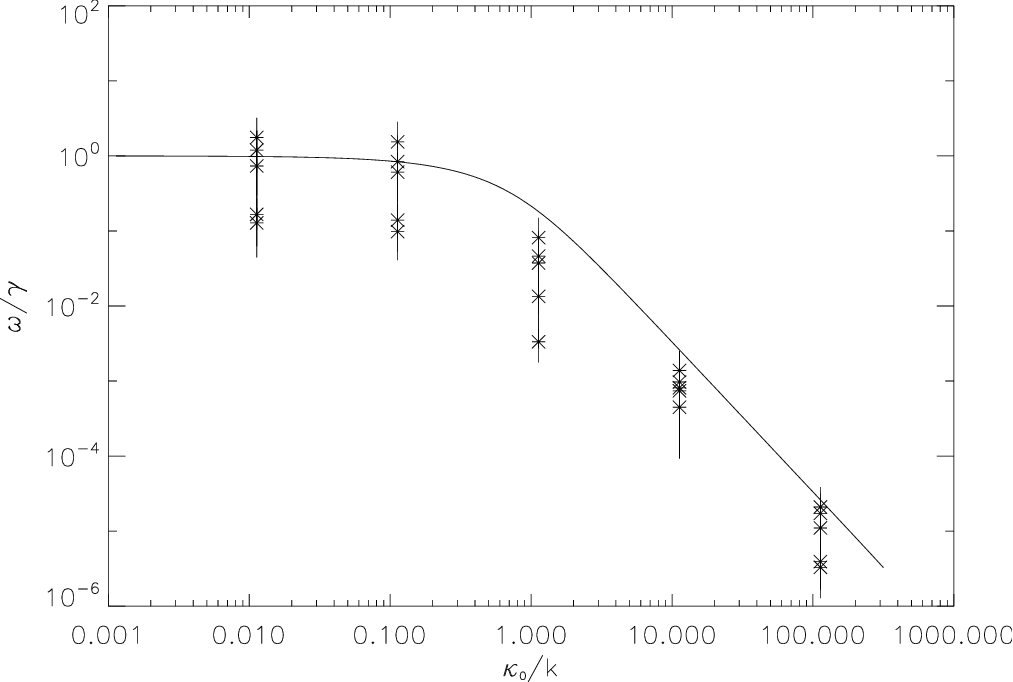} \\
\end{array}$
\caption{The dispersion relation \(\omega\) for the Spiegel Test. The left panel shows the data for polytropic cooling only, the right hand panel shows the data for the hybrid method. \label{fig:sp_omega}}
\end{center}
\end{figure*}

\noindent The key analytical result is the dispersion relation \(\omega(k)\), which is shown as the solid lines in Figure \ref{fig:sp_omega}. The points in each panel  are obtained by calculating \(\omega/\gamma\) individually for all \(2 \times 10^5\) SPH particles using equations (\ref{eq:spT}) \& (\ref{eq:spgamma}), and calculating the mean.  This is done for five separate instants in the simulation, corresponding to maximum temperatures of 10.14 K, 10.13 K 10.1 K, 10.05 K and 10.02 K respectively, and plotted for several runs with different cloud opacities (i.e. different values of \(\kappa_0/k\)).  Error bars for these points indicate the sample standard deviation.  The left panel shows the results using polytropic cooling only; the right hand panel shows the results using the hybrid method.  In the optically thin regime (low \(\kappa_0/k\)) both methods deliver the same results.  As the optical depth increases, the hybrid method approximates the curve better, as it can model the local radiation transport that occurs in the optically thick limit.  However, both methods underestimate the analytical value of \(\omega\), reflecting their approximate nature.\\

\begin{figure*}
\begin{center}$
\begin{array}{cc}
\includegraphics[scale=0.5]{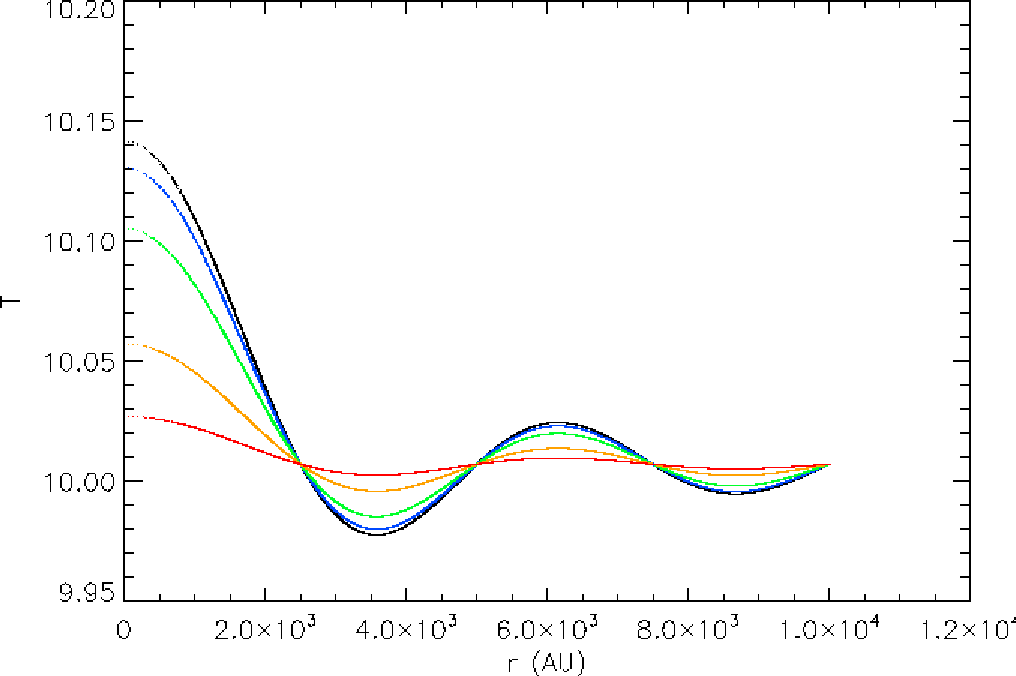} &
\includegraphics[scale=0.5]{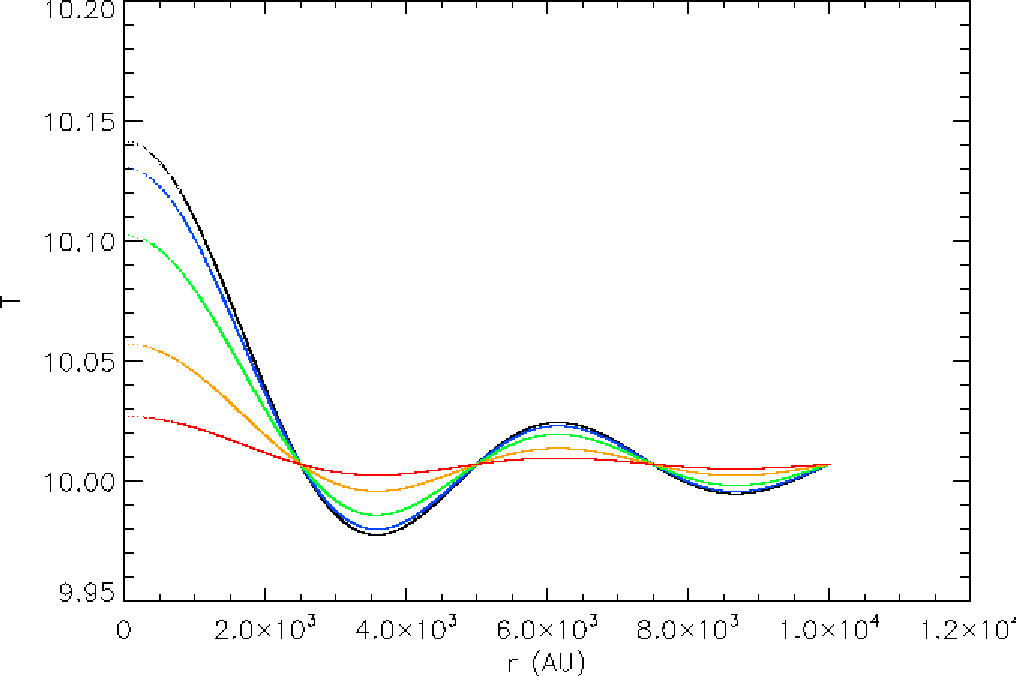} \\
\end{array}$
\caption{Temperature profiles for the thermal relaxation of an optically thin sphere ($\kappa_0/k = 0.01$).  The left panel shows the data for polytropic cooling only, the right hand panel shows the data for the hybrid method.   \label{fig:thinprofiles}}
\end{center}
\end{figure*}

\begin{figure*}
\begin{center}$
\begin{array}{cc}
\includegraphics[scale=0.5]{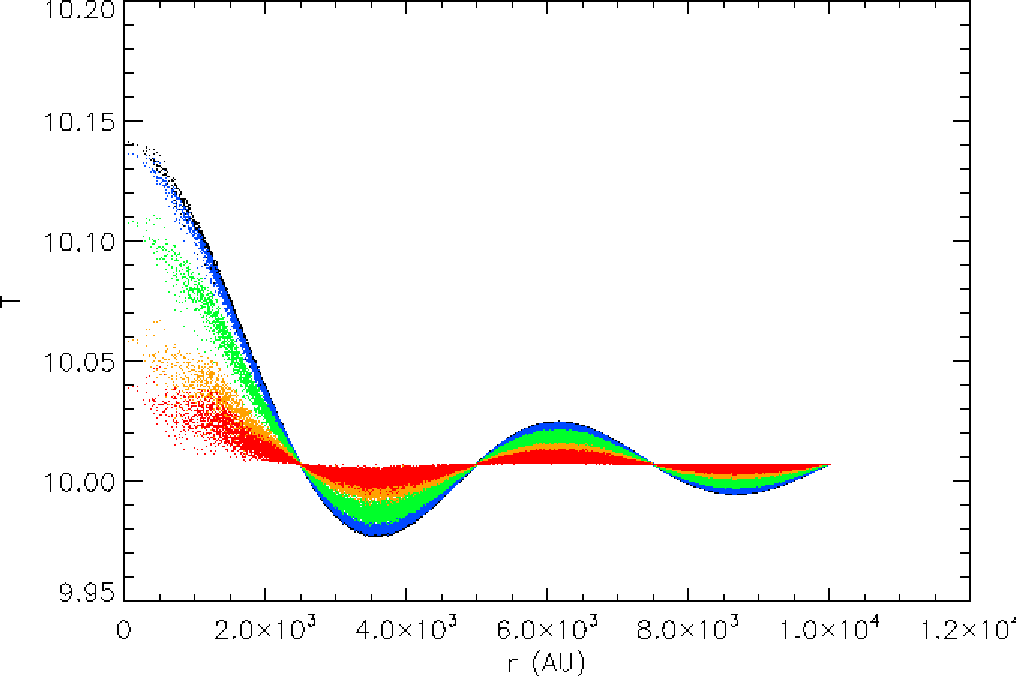} &
\includegraphics[scale=0.5]{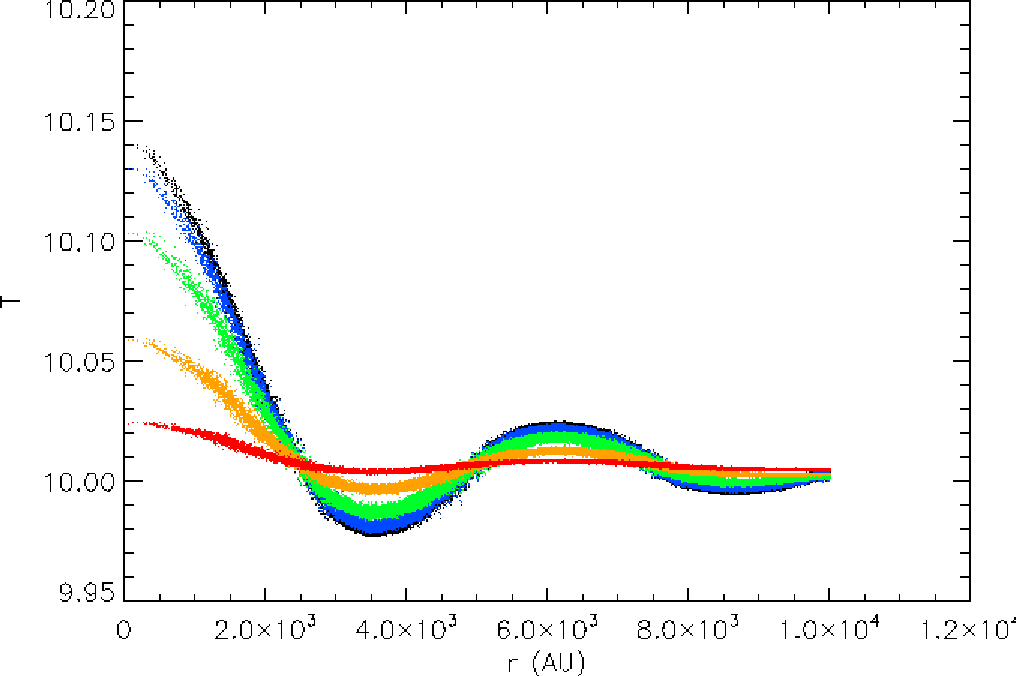} \\
\end{array}$
\caption{Temperature profiles for the thermal relaxation of an optically thick sphere ($\kappa_0/k = 100$).  The left panel shows the data for polytropic cooling only, the right hand panel shows the data for the hybrid method.  \label{fig:thickprofiles}}
\end{center}
\end{figure*}

\noindent For extra comparison, the temperature profiles of the cloud for polytropic cooling and the hybrid method are shown in Figures \ref{fig:thinprofiles} \& \ref{fig:thickprofiles}.  In the optically thin case (Figure \ref{fig:thinprofiles}), the two panels are basically identical, since flux-limited diffusion is not active in this limit; both illustrate the decaying sinusoidal function described in equation (\ref{eq:spT}).  In the optically thick case (Figure \ref{fig:thickprofiles}), the curve for polytropic cooling begins to spread, filling the regions between the troughs/peaks and 10 K.  The same panel for the hybrid method shows less spreading, retaining a more robust sinusoidal pattern.

\section{Conclusions }\label{sec:Conclusions}

\noindent This paper has presented a new means of modelling radiative transfer in SPH by fusing two well tested methods, polytropic cooling and flux-limited diffusion, in order that they may complement each other, and perform the functions that the other cannot.  By this fusion, the physics of three-dimensional frequency-averaged radiative transfer is captured without the need for complex boundary conditions, photosphere mapping or extra parameters.  Temperatures and opacities are obtained using a non-trivial equation of state which captures the effects of H$_{2}$ dissociation, H$^{0}$ ionisation, He$^{0}$ and He$^{+}$ ionisation, ice evaporation, dust sublimation, molecular absorption, bound-free and free-free transitions and electron scattering.  This data is tabulated pre-simulation for interpolation by the code. \\

\noindent The algorithm is fast: only a 6\% increase in CPU time is incurred in comparison to standard SPH simulations performed with a barotropic equation of state.  It has shown itself to be accurate in the tests outlined in the previous section: the evolution of a protoplanetary disc (with parameters proposed by Mej\'{i}a, Boley, Cai et al \citep{Mejia_1,Mejia_2}) from a uniform state through ring formation and contraction to instability; the complex thermal history of a collapsing molecular cloud (as studied by \citet{Masunaga_1}); and the smoothing of temperature fluctuations in a homogeneous, static sphere \citep{Spiegel,Masu_98}.  However, the scheme is still approximate, and can only partially describe radiative effects that occur over midrange distances (unlike the scheme proposed by \citet{BDNL}, albeit in the vertical direction only).\\

\noindent Comparisons with simulations using polytropic cooling alone have shown that the hybrid method is in effect only a small correction to the polytropic cooling method, which however can become important in some problems where temperature gradients and system geometries become complex (for example the protoplanetary disc simulations described in this paper).  \\

\noindent Future work will see this algorithm applied to a variety of protostellar and protoplanetary environments, primarily a study of initial conditions for disc formation and evolution, as well as the effects of interactions between discs and binary companions.

\section*{Acknowledgments}

\noindent The authors would like to acknowledge the referee R. Durisen for his comments that greatly improved the original manuscript.  Density plots were produced using SPLASH \citep{Price}.  All simulations were performed using high performance computing funded by the Scottish Universities Physics Alliance (SUPA).  DS and AW gratefully acknowledge the support of an STFC rolling grant (PP/E000967/1) and a Marie Curie Research Training Network (MRTN-CT2006-035890).

\appendix

\label{lastpage}

\end{document}